\documentclass[pre,amsmath,amssymb, twocolumn, showpacs]{revtex4}
\usepackage{graphics}
\usepackage{graphicx}
\graphicspath{{Figuras/}}
\usepackage{dcolumn}
\usepackage{bm}

\usepackage[utf8]{inputenc}
\usepackage[T1]{fontenc}
\usepackage{mathptmx}

\draft 

\begin{document}

\preprint{AIP/123-QED}

\title{Dynamic neighbors: \\ a proposal of a tool to characterize phase transitions}

\author{L. Aramis de Icaza Astiz}
\email{aramisdeicaza@ciencias.unam.mx}
\author{Atahualpa S. Kraemer}%
 \email{ata.kraemer@ciencias.unam.mx}
\affiliation{Departamento de F\'isica, Facultad de Ciencias, Universidad Nacional
Aut\'onoma de M\'exico,
Ciudad Universitaria, M\'exico D.F.\ 04510, M{e}xico
}%

\author{Gerardo Odriozola}%
 \email{godriozo@gmail.com}
\affiliation{\'Area de F\'isica de Procesos Irreversibles, Divisi\'on de Ciencias B\'asicas e Ingenier\'ia, Universidad Aut\'onoma Metropolitana-Azcapotzalco, Av. San Pablo 180, 02200 M\'exico, D.F., M{e}xico\\
}%

\author{Mariano L\'{o}pez de Haro}%
 \email{malopez@unam.mx}
\affiliation{Instituto de Energ\'{\i}as Renovables, Universidad Nacional Aut\'onoma de M\'exico (U.N.A.M.),
Temixco, Morelos 62580, M{e}xico.\\
}%

\date{\today}

\begin{abstract}
For molecular dynamics simulations of hard particles, we define dynamic neighbors as the distinct particles that collide with a given reference one during a specific time interval. This definition allows us to determine the distribution of the number of dynamic neighbors, its average, and its standard deviation. We will show that regardless of the time window used to identify dynamic neighbors, their distribution is correlated with diffusion coefficients, structure, and configurational entropy. Thus, it is likely that the distribution of the number of dynamic neighbors may be employed as another tool to gain insights into the dynamic behavior of hard systems. We tested this approach on 2D and 3D systems consisting of monodisperse and binary mixtures of hard disks and spheres. Results show that implementing dynamic neighbors to define order parameters can sharpen the signals where transitions take place.  

\end{abstract}

\date{\today}



\maketitle

\section{Introduction}

The study of fluids composed of particles with hardcore interactions has provided insights into the mechanisms underlying phase transitions~\cite{alder1962phase}, including transitions to quasicrystalline systems~\cite{haji2009disordered,wang2021binary}. Moreover, it has contributed to a better understanding of the jamming transition~\cite{berthierbiroli,biroli2007jamming}. 
Although hard models may appear simple, their phase behavior is strongly influenced by factors such as dimensionality of space, confinement, shape, and degree of polydispersity, resulting in remarkably rich phase diagrams \cite{royall2023colloidal}. 
For instance, in three dimensions (3D), monodisperse spheres exhibit a first-order fluid-solid transition~\cite{hoover1968melting,noya2008determination,robles2014note}. However, in two dimensions (2D), squares follow the Kosterlitz-Thouless-Halperin-Nelson-Young (KTHNY) two-step continuous melting mechanism~\cite{kosterlitz1973ordering, halperin1978theory, young1979melting}, where a tetratic phase appears in-between the isotropic-fluid and the solid phase ~\cite{donev2006tetratic,Anderson2017}, while disks undergo a first-order isotropic-hexatic fluid-fluid transition followed by a higher-order hexatic-solid transition~\cite{binder2002liquid, engel2013hard, bernard2011two, PhysRevE.73.065104, PhysRevLett.118.158001, huang2020melting}. Indeed, in 2D, several solid phases melt following complicated paths~\cite{Anderson2017}, which can involve even more than one $i$-atic phase, as the one found for superdisks~\cite{gurin2020three}. Additionally, the shape of particles, whether concave~\cite{avendano2017packing,gonzalez2021phase,qiao2023shape,jiao2008optimal,Ramirez2023densest} or convex~\cite{veerman1990phase, martinez2005effect,bautista2014phase,Anderson2017,aliabadi2018ordering,allen2019molecular,cinacchi2019hard,lopes2021phase,nasirimoghadam2022biaxial}, plays a crucial role, and the effects of confinement~\cite{fortini2006phase,basurto2021anisotropy,gnidovec2022dense,gurin2022anomalous,aliabadi2023study} can even be counter-intuitive~\cite{gurin2021enhanced}. Although some maximal packing structures have been recently established for specific high dimensions~\cite{viazovska2017sphere,cohn2017sphere}, our knowledge regarding the fluid's transition towards them remains limited~\cite{bauerschmidt2019dislocation}. 

In general, the addition of a certain degree of polydispersity tends to hinder crystallization. Polydispersity can be introduced in terms of shape or size, following various distributions such as Gaussian, binary, etc. Therefore, the effect of polydispersity can be quite complicated and challenging to generalize. However, certain distributions are known to prevent crystallization ~\cite{torquatobinary,callaham2017population,torquato2010jammed,odriozola2011equilibrium}, and there are relatively simple rules that can be applied to map the equation of state (EOS) of the polydisperse system to its monodisperse counterpart~\cite{santos2014simple,santos2023heuristic}. When crystallization is avoided, the behavior of the high-density EOS becomes protocol-dependent~\cite{torquato2010jammed}, meaning that the compression rate influences the final pressure. Faster compression rates lead to higher final pressures for a given density.

Both dynamic and structural properties typically experience notable changes during phase transitions~\cite{frenkel1994simulation}. When a fluid undergoes a transition towards a solid phase, structural modifications occur, characterized by the establishment of translational and rotational symmetries. These changes, in turn, lead to an increase in positional and bond-orientational correlations. In two dimensions (2D), fluid phases exhibit short-range correlations, while solid phases exhibit positional quasi-long-range and bond-orientational long-range correlations. In the intermediate $i$-atic phases, bond-orientational quasi-long-range correlations are preserved, but translational order is broken due to an increased number of dislocations. 
In usual phase transitions, these correlations can also be captured by single numbers known as global order parameters~\cite{zinnjustin}. These parameters are defined to have a zero value for fluids in the thermodynamic limit and ideally a value of one for a perfect crystal structure. However, during a jamming transition, global structural properties do not show significant changes, resulting in modest and continuous variations in space correlations and order parameters. This contrasts with local order parameters, which have been suggested to be useful to characterize the jamming transition~\cite{torquatobinary}. Finally, viscosity increases and diffusion coefficients decrease along fluid-solid or jamming transitions. Therefore, a combination of structural and dynamic measurements can help differentiate between fluid-solid and fluid-jammed transitions as well as provide further insights. This could shed light on some open problems in hard-sphere systems like those described in  
\cite{royall2023colloidal}.

Following this idea, we will here introduce a dynamic-based quantity that can effectively capture the processes of solidification and glassy states. We define the number of \emph{dynamic neighbors} of a particle $i$, denoted as $Z_{di}$, which represents the count of distinct particles that collide with a reference one within a given time window. Firstly, we will demonstrate that the average value of $Z_{di}$ ($\langle Z_{d} \rangle$) is correlated with the diffusion coefficient of the particles from moderate to large densities. Secondly, we will point out that the inverse of the average $\langle Z_{d} \rangle$, serves as a measure of the binding strength denoted as $\mu_Z$ and exhibits similar behavior to global order parameters during the transition from the fluid to solid phases. Moreover, $\mu_Z$ also exhibits a sudden increase when the system forms glassy states. Finally, we compute the Shannon entropy of the $Z_d$ distribution, defined as $S_{Z_d}=-\sum_{Z_d} p(Z_d)\ln(p(Z_d))$, and compare its behavior to the configurational entropy, known as $S_c$, revealing certain similarities.

The paper is organized as follows. In the next section, we provide the necessary computational and other details for our subsequent developments. This is followed by Sect.\ref{sec:quantities} in which the definition of dynamic neighbors is related to diffusion coefficients, binding strength, and Shannon entropy and illustrated for both two-dimensional and three-dimensional systems of hard particles. The paper is closed in the final section with relevant concluding remarks.

\section{Systems, order parameters, simulation details, and dynamic neighbors}
\label{sec:system and quantities}
We begin by pointing out that while the concepts discussed in this work can be applied to hard particles of any shape, degree of polydispersity, space dimensionality, and confinement, our focus will be on systems composed of monodisperse hard disks, spheres, and a binary mixture engineered to impede crystallization. 

We investigate the following systems: (i) 2D monodisperse disks, (ii) a binary mixture of disks with radii of 1 and $\sqrt{2} \approx 1.4$, in a ratio of 2:1, and (iii) 3D monodisperse spheres. To minimize the size effects, periodic boundary conditions are applied across all cases. Specifically, we employ a total of 10044 disks for the 2D scenarios and 2048 spheres for the 3D scenario. As is customary, we define the packing fraction, $\phi$, as the ratio of occupied volume to total volume. Note that during the paper, to make the figures easier to read, we plot most of the times in blueish colors the results corresponding to $2$-dimensional systems and reddish colors to $3$-dimensional systems unless otherwise stated. We also put vertical dashed lines at packing fractions corresponding to the fluid-solid transition in $3$D, fluid-hexatic, and hexatic-solid transitions in $2$D. 

Our simulations utilize the Lubachevsky–Stillinger algorithm~\cite{lubachevsky} to compress the systems until reaching a target packing fraction ($\phi$). Subsequently, we employ a molecular dynamics algorithm~\cite{sigurgeirsson} to evolve the system. Both codes are written in Julia~\cite{bezanson2017julia}. To maintain a constant temperature, we employ the velocity rescale algorithm~\cite{bussi2007canonical} in the initial step. In the subsequent step, we sample from the microcanonical ensemble. 

We used various protocols to compress our systems to a desired packing fraction $\phi$. In 2D, we compressed at a rate of $r^2 / (ct)^2$, where $c = 0.008$. For 3D systems, we used a compression rate of $r^3 / (ct)^3$, with $c = 0.001$ for slow compression and $c = 0.1$ for fast compression. The values of $c$ represent the constant growth rate of particles in the Lubachevsky–Stillinger algorithm~\cite{lubachevsky}, and $r$ is the particle radius. After reaching the desired packing fraction, we took a sample of a copy of each system while it relaxed for $\delta t = 1000$. We then let the original systems relax for $9\times 10^8$ collisions in 2D and $3.8 \times 10^7$ collisions in 3D. Finally, we took another sample after the systems had relaxed for $\delta t = 1000$. 

Length units are defined in terms of the radius of the smallest particle, $r_0$, while time units are expressed as $(\beta m)^{1/2} r_0$, where $\beta=1/k_BT$ is the reciprocal of the thermal energy and $m = 1$ denotes the mass of all particles regardless of their size. In this way, the diffusion coefficients are given in dimensionless units. 

In 2D, it is common to define an order parameter $\Psi_n$ as:
\begin{equation}
    \Psi_n = \frac{1}{N} \sum_{i = 1}^N \left | \frac{1}{Z_{i}} \sum_{j = 1}^{Z_{i}} e^{\sqrt{-1} \cdot n \cdot\theta_{ij}} \right |,
    \label{eq: order parameter}
\end{equation}

\noindent where $Z_{i}$ represents the number of neighbors of particle $i$, and $\theta_{ij}$ denotes the angle formed between the segment joining particles $i$ and $j$ and a fixed reference direction. The value of $n$ in this definition is typically set to 6 for disks (to capture a six-fold symmetry) and 4 for squares (to capture a four-fold symmetry). However, $n$ can take other values. 

Similarly, for 3D systems, the most common choice for spherical particles is~\cite{torquato2010jammed}: 
\begin{equation}
Q_6= \left(4\pi/13 \sum_{m=-6}^{m=6}| \langle Y_{6m}(\theta_{ij}, \phi_{ij}) \rangle|^2 \right)^{1/2},
\end{equation}
where $Y_{6m}(\theta_{ij}, \phi_{ij})$ represents the spherical harmonic with polar angles $\theta_{ij}$ and $\phi_{ij}$, defined by the neighbors $i$ and $j$, measured from a fixed reference frame. The ensemble average is denoted by $\langle \cdot \rangle$.

Note that these order parameter definitions do not specify the criteria for identifying neighboring particles. Consequently, it is necessary to employ a criterion to compute them. There are several methods (see \cite{van2012parameter} for a discussion about nearest neighbor definitions) but the three most popular approaches are:
\begin{itemize}
\item Simple cutoff: This method involves selecting a distance close to the first minimum of the radial distribution function. 
\item Fixing the number of neighbors, $n$: In this approach, the closest neighbors are determined by their distances, and the closest $n$ particles are selected.
\item Voronoi tessellation: This technique utilizes the Voronoi diagram, a spatial partitioning method, to determine neighboring particles based on their spatial proximity.    
\end{itemize}

Now we turn to examine all the above criteria.
When analyzing a static structure, a specific distance, $r$, can be employed to set a cutoff. In this case, the order parameters $\Psi_n$ and $Q_6$ become functions of $r$. If $r$ is chosen to be too small, $\Psi_n(r)$ becomes highly noisy and may be undefined when the number of neighbors approaches zero. On the other hand, if $r$ is too large, $\Psi_n(r)$ approaches zero. In the case of a crystalline system, it is relatively clear which particles should be considered close neighbors. However, for systems lacking translational symmetry, polydisperse systems, or unsymmetrical particles, selecting a unique distance to define neighbors becomes more challenging. It is worth mentioning that in this direction a method has been proposed  to choose the right distance using purely geometrical information as with the Voronoi cell \cite{van2012parameter}. However, this still does not solve the problem of polydisperse systems or unsymmetrical particles. 

Alternatively, one can fix the number of neighbors by considering the first $n$ closest particles as neighbors~\cite{haggstrom1996nearest}. Choosing the value of $n$ can be useful when there is prior knowledge of a reference structure, such as the maximum packing fraction array, where neighbors are defined based on contact. For example, in the case of 2D squares, $n$ would be set to 4, for 2D disks $n$ would be 6, and for 3D spheres, $n$ should be 12, as 12 represents the number of contact neighbors in face-centered cubic and hexagonal close-packed structures. However, for particles with a rhombus shape, the choice is not as straightforward, as particles touching the corners may have center-to-center distances shorter than those touching the edges. Similar considerations apply to other shapes such as tetrahedra~\cite{haji2009disordered}, rounded squares~\cite{avendano2012phase}, superdisks~\cite{jiao2008optimal,gurin2020three}, superspheres~\cite{maher2022characterization}, superellipses~\cite{torres2022hard}, and rounded rectangles~\cite{martinez2022effect}. Similar challenges arise when studying mixtures of spheres~\cite{yi2012radical}, where the maximum packing fraction array may exhibit different numbers of neighbors for each species. Confinement effects can also lead to variations in the number of neighbors as a function of position, as particles are confined by cylindrical cavities~\cite{de2014domain,jin2020shape}.

Finally, one of the most elegant alternatives is the use of Voronoi cells~\cite{voronoi1,voronoi2} to define neighboring particles based on the sharing of edges in 2D (or faces in 3D) of their cells. Voronoi cells have various applications, ranging from social sciences~\cite{laver2011party} to biology~\cite{bock2010generalized}, and are frequently employed to analyze the topological properties of random close packing~\cite{coxeter1958close, voronoi1}. One key advantage is that Voronoi cells are well-defined regardless of the shape, degree of polydispersity (in shape or size), confinement, or spatial dimensionality. Voronoi cells have also been utilized in estimating free volume to deduce the Vogel-Fulcher-Tamman law~\cite{elliott1983physics, wales2003energy}, as well as in assessing configurational entropy \cite{kumar2005voronoi}. Moreover, they have served as the foundation for cage theory, used to estimate densities at which phase transitions occur~\cite{kraemer2008use}, and have been directly applied to measure the transition between a glassy state and the crystalline phase in hard-sphere systems~\cite{jin2010first}. 

However, it is important to note that while Voronoi cells are useful in describing phase transitions, they do not guarantee that neighboring particles are the closest or would produce direct contacts when compressed to achieve the maximum packing fraction. Therefore, the definition of a Voronoi neighbor does not reflect the true particle interactions, and it can be computationally more expensive than alternative methods~\cite{engel2013hard}, particularly when dealing with non-spherical particle shapes~\cite{schaller2013set}. The computational complexity increases rapidly as the dimension of the system grows~\cite{boissonnat1998voronoi}. Despite these considerations, we are employing this approach to compare the results obtained from dynamic neighbors with those obtained from order parameters. 

\begin{figure}
    \begin{center}
    \includegraphics[width=245pt]{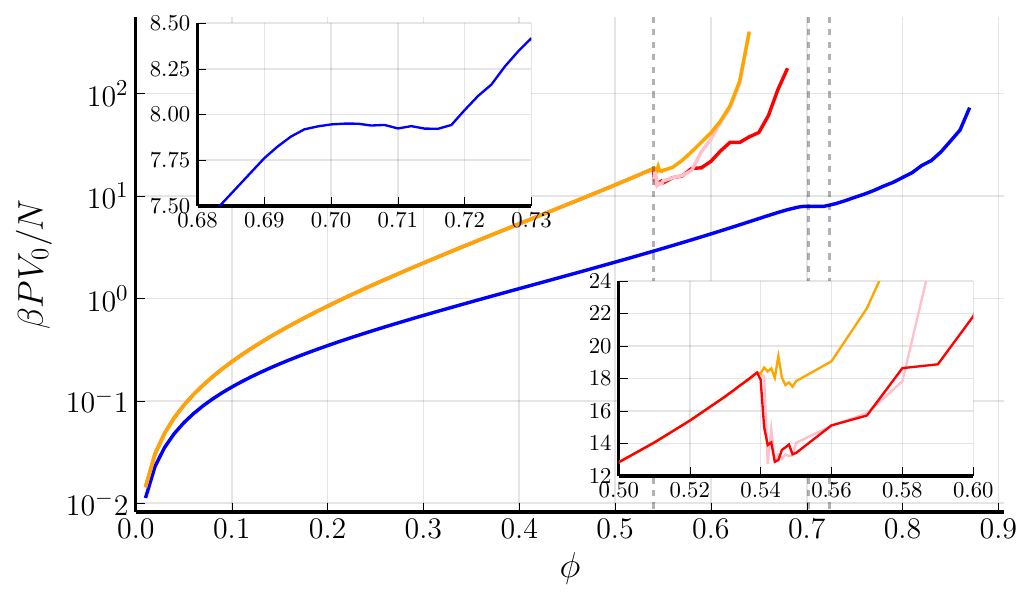}
    \end{center}
    \caption{(Color online) EOS for monodisperse systems calculated by means of equation (\ref{eq:EOS}) as a function of $\phi$, for hard disks (dark blue line) and hard spheres (reddish lines). The dark red, orange, and pink lines correspond to the slow compression of spheres, fast compression with a short sampling time, and fast compression with a large sampling time, respectively. The dashed vertical lines define the regions corresponding to transitions. The insets correspond to a zoom of the region where the transitions occur.The vertical dashed lines correspond to specific values of $\phi$, namely $\phi =$ $0.54$, $0.702$, and $0.724$.}
    \label{fig:colisions}
\end{figure}

While previous neighbor definitions rely on a given static configuration, we propose a different approach based on collision dynamics. The total number of collisions, denoted as $N_{\text{coll}}$, is related to the pressure through the equation~\cite{woodcock1997computation, engel2013hard, li2022hard}
\begin{equation}
\label{eq:EOS}
    \frac{\beta PV}{N} = 1+\frac{2r  N_{\text{coll}}\sqrt{\pi m\beta }}{N \delta t},
\end{equation}
where $m$ is the mass of the particle,  $\delta t$ represents the time window during which the collisions occur, and $r$ is the particle radius. In Figure~\ref{fig:colisions}, we present the pressure as a function of $\phi$ calculated with the total number of collisions. We have chosen the same form as \cite{alder1962phase} or \cite{li2022hard} $\beta P V_0 /N$, where $V_0 = N 8 \sqrt(3) r^2$ for two-dimensional systems, and $V_0 = 24 \sqrt{2} N r^3$ for three dimensional systems. 
The upper and lower insets correspond to zoomed views around the phase transition for the 3D and 2D cases, respectively.   
In all cases, we observe that the behavior
is similar to the pressure behavior described elswhere~\cite{alder1962phase, engel2013hard, li2022hard}. This happens regardless equilibrium is set (dark blue and red lines) or not (orange, and pink lines). Hence, it seems reasonable to utilize the dynamic information from collisions to define neighbors, rather than restricting to static information. 

We define the number of \emph{dynamic neighbors} of a particle $i$, denoted as $Z_{di}$, as the number of different particles that collide with $i$ within a time window $\delta t$. Consequently, both $Z_{di}$ and the average number of dynamic neighbors, $\langle Z_d \rangle= \tfrac{1}{N} \sum_i Z_{di}$, depend on $\delta t$. In a sufficiently large $\delta t$ and a finite fluid system, each particle collides with all the others. Conversely, for a small enough $\delta t$, there will be no collisions. Hence, it is convenient to define an \emph{adequate} time window that is large enough for all particles to have multiple collisions with others (not necessarily different particles), but not so large that most particles have collided with all the others.

To choose the appropriate value of $\delta t$, let us first note that unlike neighbors defined using Voronoi cells or a cutoff, $Z_{di}$ can be significantly large. We can leverage this distinction for three reasons: (i) it allows us to obtain a broader and more detailed neighbor distribution, particularly useful at middle and low densities where valuable information can be extracted, (ii) it results in a sharper distinction between global order parameters for fluid and solid (or arrested) states, and (iii) we can introduce a \emph{binding parameter}: $\mu_Z \propto \tfrac{Z_{min}}{\langle Z_{d} \rangle }$ which quantifies the strength of the entropic bonds~\cite{Harper19}, where $Z_{min}$ is a normalization parameter.

\begin{figure}
    \includegraphics[width=245pt]{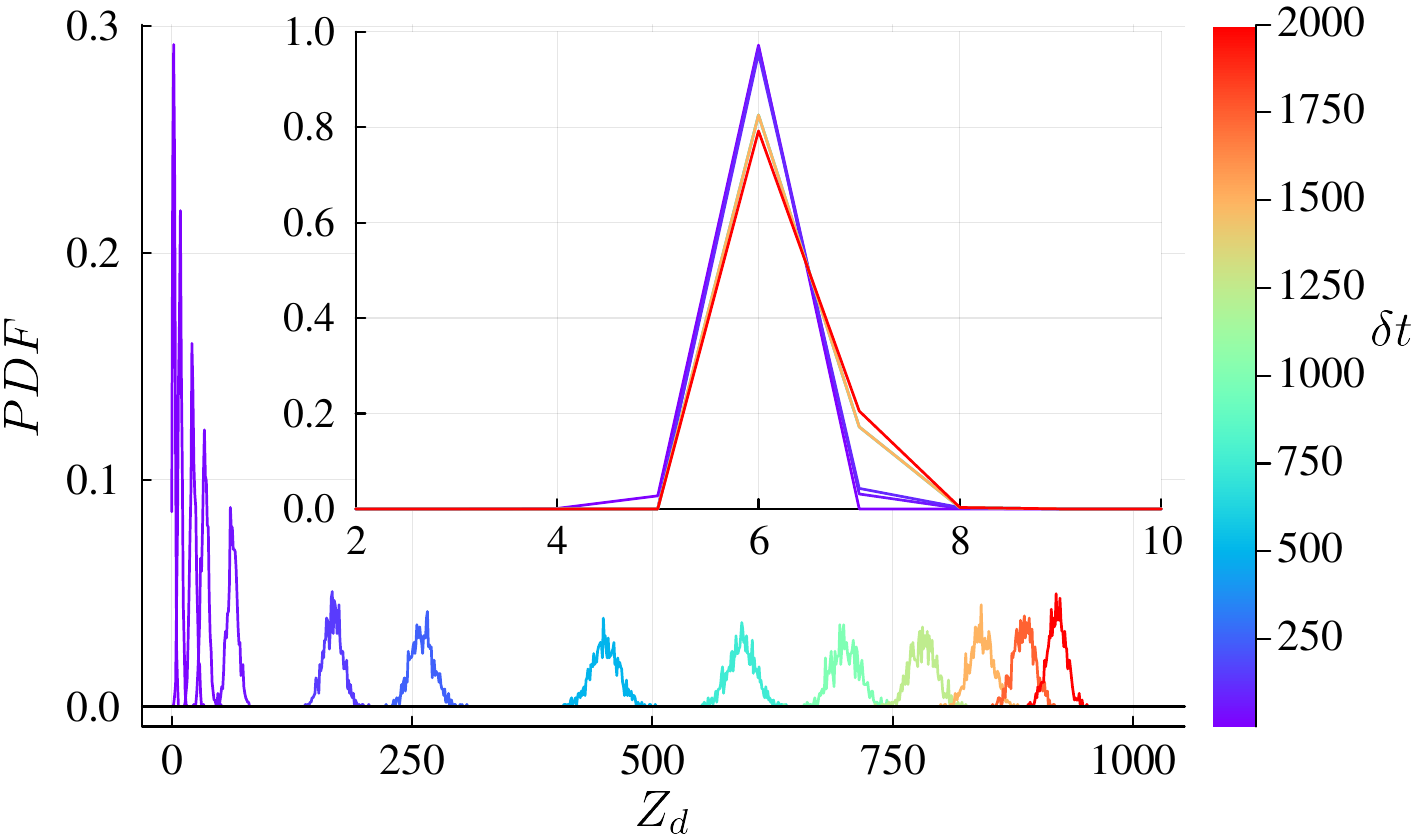}
    \caption{(Color online) Probability density functions (PDFs) of the dynamic neighbors, $Z_{d}$, for a two-dimensional system with 1024 particles at $\phi = 0.2$, and $\phi = 0.8$ in the inset. Different $\delta t$ values are represented by the colorbar. }
    \label{fig: distributions_example}
\end{figure}

Figure \ref{fig: distributions_example} illustrates the changes in the neighbor distribution as a function of $\delta t$ for hard disks at $\phi = 0.2$ (main panel) and $\phi = 0.8$ (inset). In the fluid region, larger $\delta t$ values lead to higher $\langle Z_{d} \rangle$ values and a wider distribution of $Z_{di}$. In contrast, arrested or solid systems exhibit distributions that remain almost constant across varying $\delta t$, as depicted in the inset of Fig.~\ref{fig: distributions_example}. Note that because the distributions are almost symmetric, the most likely $Z_{di}$ and $\langle Z_{d} \rangle$ almost match in the limit of infinite sampling. In practice, there is less variance in the most likely value than in $\langle Z_{d} \rangle$, so in simulations we use the most likely $Z_{di}$ as a measure of $\langle Z_{d} \rangle$ except for $\mu_{Z}$, where the inverse of $\langle Z_d \rangle$ amplified the difference between $\langle Z_{d} \rangle$ and the most likely $Z_{di}$. However, all the measurements were made using both values, obtaining similar results. 

\begin{figure}
    \centering
   \includegraphics[width=245pt]{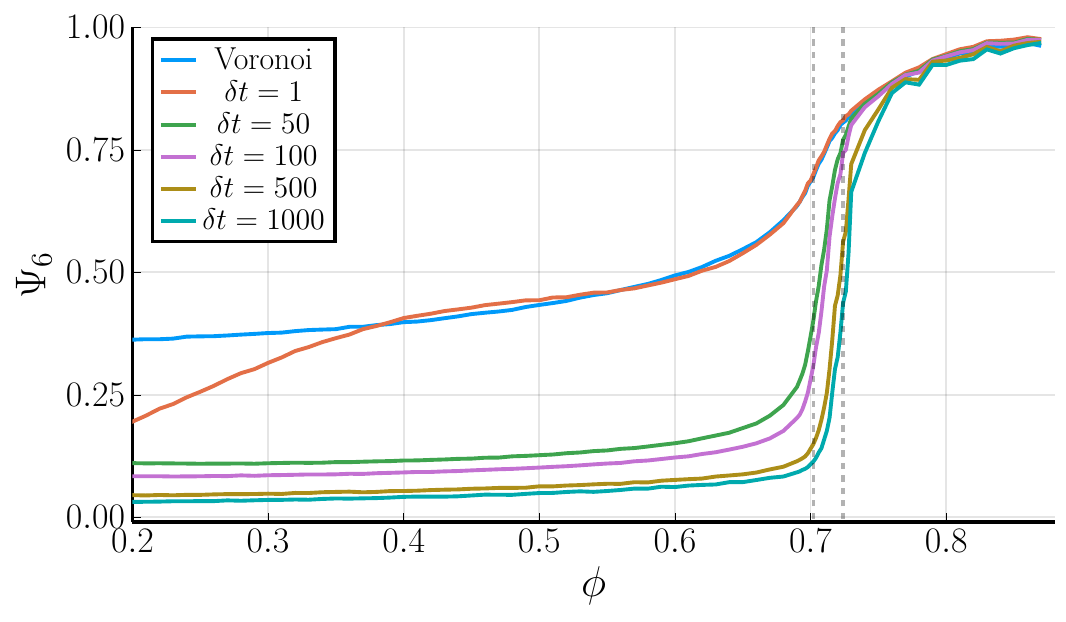}
    \caption{Bond-orientational order parameter $\Psi_6$ as a function of the packing fraction, $\phi$, computed with the Voronoi tesselation, and from dynamic neighbors with different $\delta t$ values. The vertical dashed lines correspond to specific values of $\phi$, namely $\phi =$ $0.702$ and $0.724$.}
    \label{fig: Psi6}
\end{figure}

Figure~\ref{fig: Psi6} compares the $\Psi_6$ local order parameter for disks using Voronoi tessellation and dynamic neighbors with different $\delta t$ values. As observed, both the tessellation and dynamic neighbors yield low values for the fluid phase and high values for the solid phase. However, there are some differences. $\Psi_6$ exhibits a significant decrease with increasing $\delta t$ in the fluid phase related to the fact that in such phase, the number of neighbors increases fast with $\delta t$, turning the measurement from local to global, while in the solid phase, the neighborhood keeps local and the decrease is less pronounced. Consequently, when employing dynamic neighbors, the transition between phases becomes sharper with larger $\delta t$ values compared to using Voronoi tessellation. This behavior aligns with the desirable characteristics of a well-designed order parameter~\cite{torquatoreview2018}.    

Consequently, a larger value of $\delta t$ allows for improved sampling of the $Z_{d}$ distribution while accentuating the distinctions between the fluid and arrested phases. However, in the fluid phase, increasing $\delta t$ also results in greater utilization of computational resources, such as the need to define larger lists and execute more extensive loops for computing the order parameters. Henceforth, we set $\delta t$ to $1000$.

\begin{figure}
    \centering
    \includegraphics[width=245pt]{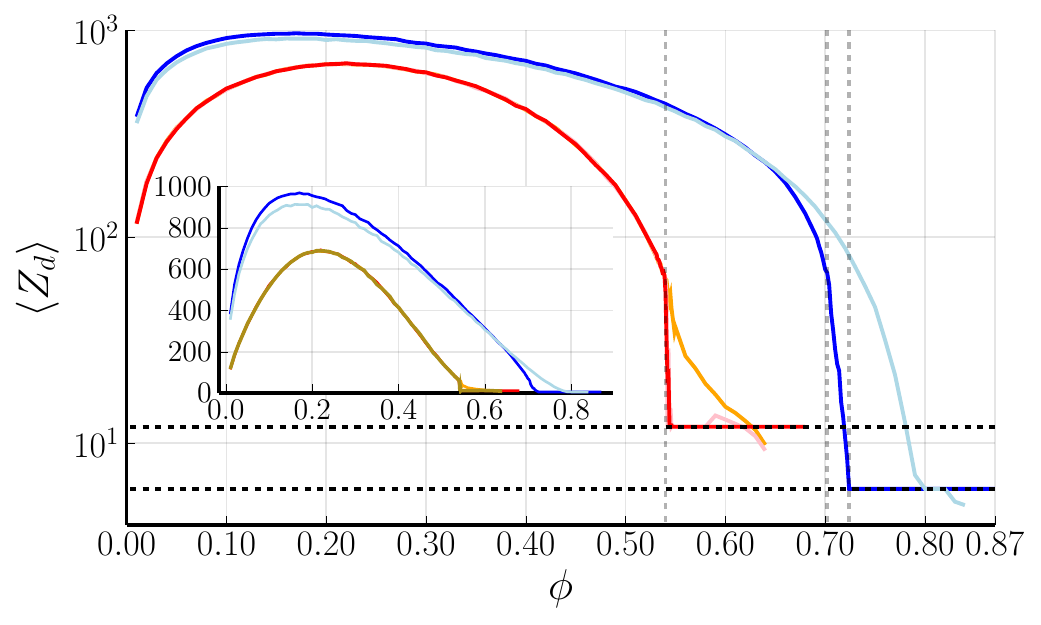}
    \caption{(Color online) $ \langle Z_d\rangle$ as a function of $\phi$, for a monodisperse system of hard disks (dark blue line), a binary mixture of hard disks (light blue line), and a monodisperse system of hard spheres (reddish lines). The dark red, orange, and pink lines correspond to the slow compression of spheres, fast compression with a short sampling time, and fast compression with a large sampling time, respectively. The dashed vertical lines define the regions corresponding to transitions. The insets show the same data but with linear axes. The horizontal dashed lines correspond to $6$ and $12$ dynamic neighbors. The vertical dashed lines signal the fluid-solid and fluid-jammed transitions. The vertical dashed lines correspond to specific values of $\phi$, namely $\phi =$ $0.54$, $0.702$, and $0.724$.}
    \label{fig:NNPF}
\end{figure}

The shape of $\langle Z_{d}\rangle(\phi)$ is shown in figure \ref{fig:NNPF} for monodisperse disks, the binary mixture of disks, and spheres, this last case when following slow and fast compression protocols. In all cases, the first thing to note is that for the ideal gas limit we get $\langle Z_{d}\rangle(0) \rightarrow 0$, since there are no collisions. As the packing fraction increases, $\langle Z_{d}\rangle(\phi)$ augments as $\phi^{1-1/d}$ due to the fact that for low packing fractions, most collisions between particles occur only once. Note that the mean free path $\langle l \rangle \propto t_{av} = \delta t N/(2N_{coll}) \propto V/(N v_1'(r)) \propto r / \phi$, where $t_{av}$ is the mean time between collisions, $V$ is the system volume, and $v_1'(r)$ is the surface of a sphere of radius $r$ ~\cite{serway2018physics}. The last relation follows since $v_1(r) \propto v_1'(r) r$, and this implies $ \langle Z_{d} \rangle \propto N_{coll}\propto \phi/r \propto \phi^{1-1/d}$. However, if the packing fraction continues to grow, collisions between the same pairs become more common, so the number of dynamic neighbors decreases. So, $\langle Z_{d} \rangle$ is $0$ for $\phi = 0$, then grows as a power law while collisions between the same pairs are scarce, then decelerate the growth of the number of neighbors as this type of collision grows in frequency, and finally reaches a maximum and starts decreasing, when the collisions involve practically always their caging neighbors. This effect turns dominant along a fluid-solid transition, where $\langle Z_{d} \rangle$ becomes $6$ or $12$ for the 2D and 3D cases, respectively. At the fluid-solid transitions, the decay of $\langle Z_{d}\rangle (\phi)$ with $\phi$ is quite steep, contrasting with the formation of glassy states, where the decay rate is slower and depends on the simulation protocol. 

\section{Using dynamic neighbors to compute different quantities}
\label{sec:quantities}
 
In this section, our focus is to examine the relationship between $\langle Z_d \rangle$ and the diffusion coefficient $D$, order parameters, and entropy. 

\subsection{Diffusion coefficients}

To estimate the diffusion coefficients at packing fractions near a phase transition. We considered the scenario where all particles except particle $i$ remained fixed. In this case, we tile the system using Voronoi cells of the fixed particles. Given the system's proximity to a phase transition, most of the time that particle $i$ visited a Voronoi cell, it resulted in a collision with the particle contained within that cell. Thus, $\langle Z_d \rangle$ serves as an approximation for the number of visited Voronoi cells. In other words, we consider the trajectory of particle $i$ as a random walk within the network formed by the Voronoi cells, and $\langle Z_d \rangle$ represents the average number of visited cells within the time window $\delta t$.

The number of visited cells has been computed for cases where the random walk occurs on a periodic (square, cubic, or hypercubic) lattice, leading to two distinct scenarios based on the dimensionality of the system \cite{dvoretzky1951some}. In a 2D system, the relationship can be expressed as follows:
\begin{equation}
\label{eq:2D_num}
\langle Z_{d} \rangle \sim \frac{\pi b_{2D} \delta t}{\log(b_{2D} \delta t)},
\end{equation}
Here $b_{2D}$ is a constant depending on the packing fraction and inversely proportional to the average time the particle $i$ spends in a Voronoi cell. Specifically, it serves as a normalization constant for the average time it takes a particle to move from one cell to a neighboring cell.

On the other hand, for three or higher-dimensional systems, the relationship is given by:
\begin{equation}
\label{eq:3D_num}
\langle Z_{d} \rangle \sim b_{3D} \delta t.
\end{equation}
In this case, $b_{3D}$ fulfills a similar role as $b_{2D}$.

We can determine $b_{2D}$ and $b_{3D}$ as functions of $\langle Z_{d} \rangle$ and $\delta t$ by inverting equations~(\ref{eq:2D_num}) and~(\ref{eq:3D_num}). The inversion for equation~(\ref{eq:3D_num}) is straightforward: $b_{3D} = \langle Z_{d} \rangle / \delta t$. However, in the case of equation~(\ref{eq:2D_num}), inversion involves using the transcendental Lambert W function, denoted as $W(x, -1)$ \cite{barry2000analytical}. The result is expressed as $b_{2D} = \exp\left(-W\left(-\tfrac{\pi}{\langle Z_{d} \rangle}, -1\right)\right)/\delta t$, which requires a numerical solution. It is worth noting that $b_{2D}$ is a complex number, but its imaginary part is zero if the system is the fluid. Nevertheless, as the diffusion coefficient approaches zero, the imaginary part grows, resulting in a finite quantity. As the imaginary part lacks a clear physical interpretation, we focus only on the real part of $\exp\left(-W\left(-\tfrac{\pi}{\langle Z_{d} \rangle}, -1\right)\right)/\delta t$ in the following analyses. 

The diffusion coefficients in a random walk on a lattice has the form $D \propto \tfrac{\xi^2}{\Delta t}$, where $\xi$ is the distance between two neighboring lattice vertices, and $\Delta t$ is the average time it takes to move from one vertex to its neighbor, proportional to $1/b_{2D}$ and $1/b_{3D}$. Hence, we obtain:
\begin{equation}
\label{eq:approxD1}
D \sim \begin{cases} a_{2D} \xi_{2D}(\phi)^2 \exp\left(-W\left(-\frac{\pi}{\langle Z_{d} \rangle}, -1\right)\right)/\delta t, & \text{for 2D} \\
a_{3D}\xi_{3D}(\phi)^2 \langle Z_{d} \rangle/\delta t, & \text{for 3D}
\end{cases} ,
\end{equation}
where $a_{2D}$ and $a_{3D}$ are constants independent of $\phi$. The next step is to calculate $\xi_{2D}(\phi)$ and $\xi_{3D}(\phi)$. For this purpose, we propose a power-law function of the form $(\phi_c -\phi)^{\beta_d}$, where $\phi_c$ represents the maximal packing fraction of the system, and $\beta_d$ is a real constant. This results in the modified equation~(\ref{eq:approxD1}):

\begin{equation}
\label{eq:approxD}
D \sim \begin{cases} a_{2D} (\phi_c -\phi)^{\beta_{2D}} \exp\left(-W\left(-\frac{\pi}{(\langle Z_{d} \rangle-6)}, -1\right)\right)/\delta t, & \text{for 2D} \\
a_{3D}(\phi_c -\phi)^{\beta_{3D}} (\langle Z_{d} \rangle-12) / \delta t , & \text{for 3D}
\end{cases} ,
\end{equation}
where the model introduces two parameters for each case: the coefficients $a_{2D}$ and $a_{3D}$, and the exponents $\beta_{2D}$ and $\beta_{3D}$. It is worth noting that, to ensure $D = 0$ when $\phi$ reaches its maximum value, we have modified $\langle Z_d \rangle$ to $\langle Z_d \rangle - Z_c$, where $Z_c = 6$ and 12 for the 2D and 3D cases, respectively. 

To validate this formula, we conducted measurements of the mean square displacement of particles in both 2D and 3D monodisperse hard-sphere fluids. Subsequently, we obtained the diffusion coefficients for each packing fraction by fitting the data using a linear function of time. To ensure the accuracy of the diffusion coefficients, we excluded short time scales from the fitting process, selected a time window that encompassed at least $10^8$ collisions, and sampled every $5 \times 10^4$ collisions. This process was repeated five times for each value of $\phi$ to average the results and obtain the final diffusion coefficients. It is worth noting that this large number of collisions prevented us from accurately determining the diffusion coefficients for metastable states, as the system typically crystallized during the measurement. Therefore, we did not compute $D$ for glassy systems. With these diffusion coefficients we fitted the parameters of equation \ref{eq:approxD} obtaining $a_{2D} = 5$, $a_{3D} = 5.2$ and $\beta_{d} = d$, where $d$ is the dimension of the system.

\begin{figure}[htb]
   \centering
    \includegraphics[width=245pt]{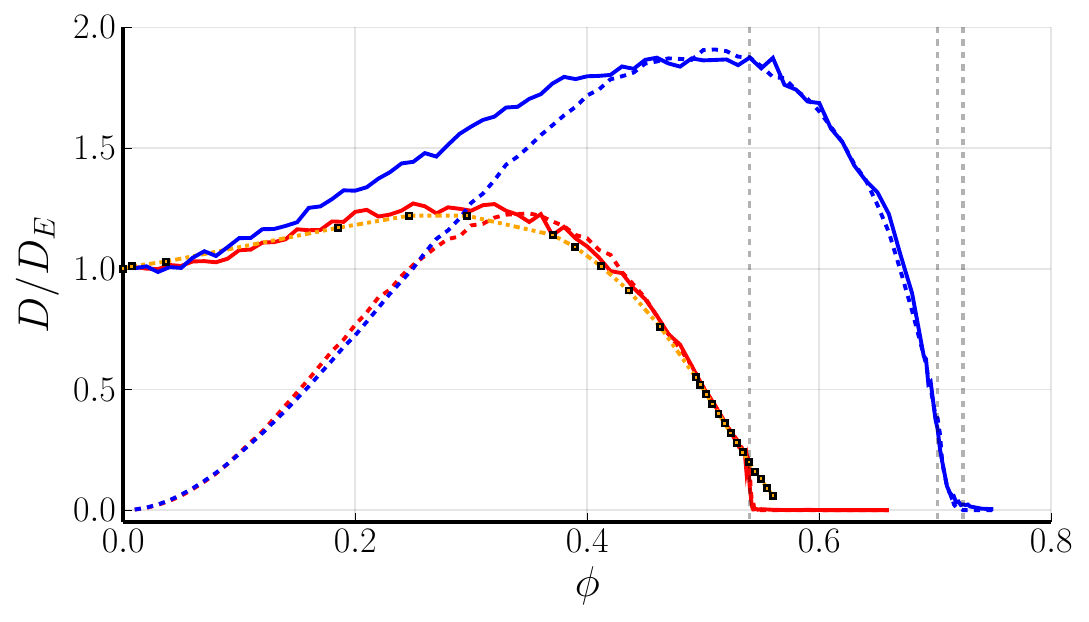}
    \caption{(Color online) Comparison of measured diffusion coefficients (solid lines) and their approximations using equation~\ref{eq:approxD} (dashed lines). The values of $a_{2D} = 5$, and $\alpha_{3D} = 5.2$ were used for the approximation. The diffusion coefficients are plotted as a function of the packing fraction $\phi$ for monodisperse systems of hard disks (blue) and hard spheres (red) at equilibrium. The yellow dashed line with squares corresponds to simulations results from Speedy~\cite{speedy1987diffusion}. The vertical dashed lines correspond to specific values of $\phi$, namely $\phi =$ $0.54$, $0.702$, and $0.724$.}
\label{fig: diffusionC}
\end{figure}

In Figure~\ref{fig: diffusionC}, we present the diffusion coefficients obtained from simulations and normalized with Enskog estimation \cite{speedy1987diffusion, sigurgeirsson2003transport, heyes2007self} alongside our approximation using $\langle Z_{d} \rangle$ as a function of the packing fraction $\phi$. Additionally, we have included the results from Speedy~\cite{speedy1987diffusion} as symbols for comparison. Remarkably, the measured diffusion coefficients show considerable agreement with our approximation. One significant advantage of employing dynamic neighbors to estimate the diffusion coefficient is its applicability to metastable systems. Since the measurements require much shorter durations, the likelihood of crystallization is minimized. This characteristic allows us to explore and analyze systems that would be challenging to study using conventional methods. Furthermore, it is worth noting that we expect Equation~(\ref{eq:approxD}) to be applicable to dimensions larger than 3, albeit with different parameter values, extending its potential use to higher-dimensional systems.

Another important observation is the difference in diffusivity between 2D systems and higher dimensions, stemming from two key factors. Firstly, in 2D, the number of dynamic neighbors does not decline abruptly as it does in 3D, which influences the overall dynamics. Secondly, the relationship between the dynamic neighbors and $b_{d}$ follows a distinct mathematical nature for each dimension, leading to different behaviors of $b_{d}$ in each case. Consequently, the average time that a particle takes to exchange its position with another one differs between 2D and higher dimensions.

\begin{figure}[htb]
   \centering
    \includegraphics[width=245pt]{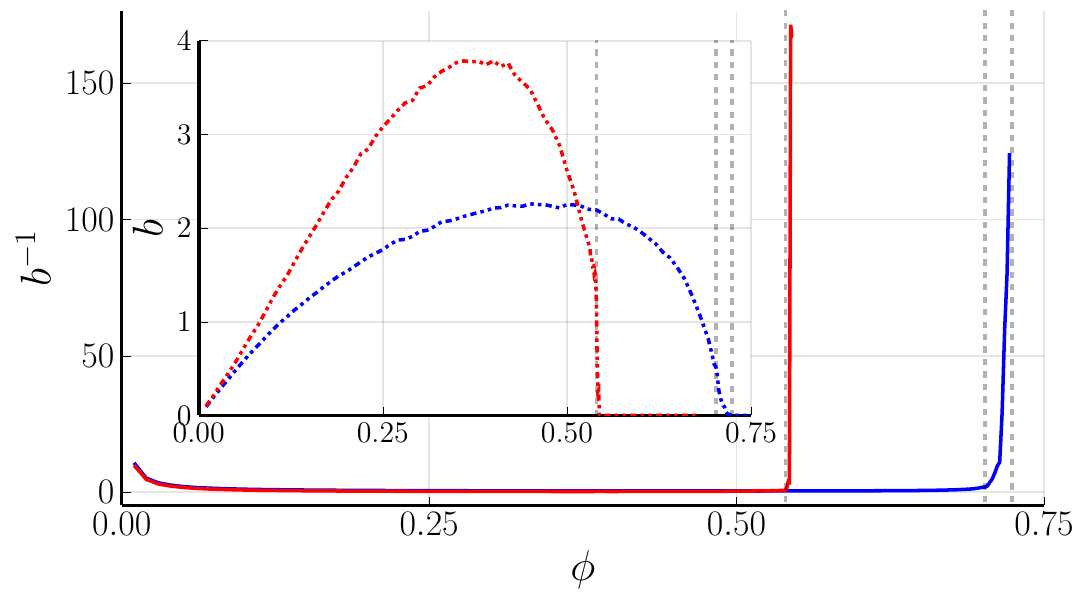}
    \caption{(Color online) Plot of parameters $1/b_{2D}$ (solid-blue) and $1/b_{3D}$ (solid-red) as functions of the packing fraction. The inset displays the data corresponding to $b_{2D}$ (dashed-blue) and $b_{3D}$ (dashed-red). Vertical lines are drawn at $\phi =$ $0.54$, $0.702$, and $0.724$.}
\label{fig:b}
\end{figure}

Finally, figure \ref{fig:b} illustrates the behavior of $1/b_{3D}$ and $1/b_{2D}$, which are proportional to the hopping time of a particle to exchange its position with a neighboring particle. We notice that, because the exchange of positions between neighboring particles is mainly due to dislocations, $1/b_{3D}$ and $1/b_{2D}$ may be useful to study the formation and diffusion of dislocations and with it, study possible formation of topological phases.
On the other hand, the contrasting trends showcased in the figure exemplify the diverse dynamics observed in 2D and higher-dimensional systems, showing an abrupt change in the three-dimensional system but a continuous increase for the hard-disc fluid. Understanding these differences is crucial for comprehending the diffusion mechanisms in complex systems and for tailoring materials with specific properties based on their dimensionality. 

\subsection{$\mu_{Z}$ and the bond-orientational order parameters}

We previously mentioned that $\mu_{Z} \propto \frac{1}{\langle Z_{d} \rangle}$ 
can be regarded as a measure of the strength of the entropic bond. Therefore, we can define $\mu_{Z} =  \frac{Z_{min}}{ \langle Z_{d} \rangle }$, where the constant $Z_{min}$ is chosen $Z_{min} = 6$ for $2$D systems and $12$ for $3$D systems. 
By doing so, we expect $\mu_{Z} \rightarrow 1$ for arrested (or solid) structures and $\mu_{Z} \rightarrow 0$ for the ideal gas limit. As stated elsewhere~\cite{zachary2009hyperuniformity}, this desired behavior is often not achieved when searching for an appropriate order parameter in various situations. However, it is important to note that $\mu_{Z}$ is not an order parameter as it does not measure the translational or rotational symmetry of configurations, nor their ensemble average. Instead, it solely reflects the ability of the structure to remain unchanged over time. Hence, it is a dynamic property.

The dependence of the bond-orientational order parameter on the value of $n$ implies that certain types of order may not be detectable in certain systems. For example, the most densely packed array would appear disordered when measuring the orientational order parameter $\Psi_5$, while a quasicrystal with pentagonal symmetry would not exhibit clear order when using $\Psi_6$. However, by utilizing $\mu_{Z}$, which is independent of the system's geometry but relies on its dynamic properties, we can overcome these limitations. When the system experiences a loss of rotational or translational symmetry, it also affects the proportion of collisions involving the same particles. A similar behavior is expected for glassy states, as the number of dynamic neighbors significantly decreases. As a result, we anticipate that $\mu_{Z}$ will exhibit abrupt changes in its value, making it a suitable indicator for detecting and characterizing such transitions. 

\begin{figure}
   \centering
    \includegraphics[width=245pt]{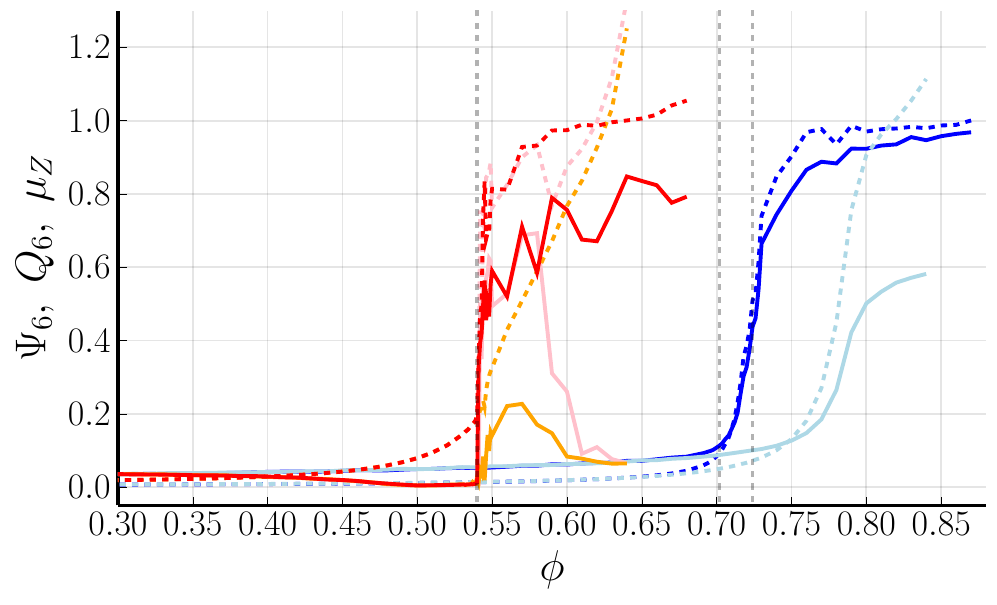} 
    \caption{(color online) Bond-orientational order parameters $\Psi_6$ and $Q_6$ (solid lines) and the binding parameter $\mu_{Z}$ (dashed lines) as a function of the packing fraction for 3D systems (red-orange lines), and 2D systems (blue lines). Red and dark blue lines correspond to systems showing a phase transition, while orange and light blue lines are used for systems showing glassy states. Pink lines correspond to the systems where the compression rate is fast, but the relaxation time is long. The vertical dashed lines correspond to specific values of $\phi$, namely $\phi =$ $0.54$, $0.702$, and $0.724$.}
\label{fig: order parameters and BF}
\end{figure}

Figure~\ref{fig: order parameters and BF} displays the orientational order parameters $\Psi_6$ (for the 2D systems) and $Q_6$ (for the 3D systems) as defined in \cite{torquato2010jammed} using dynamic neighbors, along with $\mu_{Z}$, as functions of the packing fraction $\phi$ for all the systems studied in this paper. We observe that $\Psi_6$ and $\mu_{Z}$ exhibit close agreement for monodisperse hard disks, indicating a transition around $\phi \approx 0.7$. This finding is consistent with previous observations using alternative definitions of close neighbors \cite{huerta2003role}. For the binary mixture, we again observe a similarity between $\mu_{Z}$ and $\Psi_6$, with glassy dynamics occurring at approximately $\phi \approx 0.75$ and a change of behavior at approximately $\phi \approx 0.8$.

In the case of 3D systems, $Q_6$ and $\mu_{Z}$ exhibit distinct behaviors before the phase transition. However, as the transition occurs, these parameters become more correlated, both with an abrupt change, indicating the transition at approximately $\phi \approx 0.54$. For the jammed system, $Q_6$ remains very low, suggesting a lack of structural order. Interestingly, there is a small jump in $Q_6$ prior to the appearance of glassy states. In contrast, $\mu_{Z}$ continues to increase throughout the entire packing fraction range, with a slight change in its behavior around $\phi = 0.63$, which is close to the maximum random jamming point.

We observe a close resemblance between the behavior of $\mu_{Z}$ and the bond-orientational order parameter when the latter successfully captures a phase transition. However, they may differ when the bond-orientational order parameter fails to recognize a glassy dynamic. This discrepancy can occur in systems with intricate structures, such as those involving non-spherical particles, mixtures, or confinement, or it may result from a poor choice of $n$ in the bond-orientational order parameter. In such cases, $\mu_{Z}$ offers an advantage as a dynamic property that remains unaffected by the non-trivial symmetries exhibited by a particular system. Thus, $\mu_{Z}$ provides a robust and versatile alternative order parameter that can be particularly useful in situations where the final structure is unknown or complex.

\subsection{Shannon entropy of the $Z_d$ distribution}

\begin{figure}
  \centering
    \includegraphics[width=245pt]{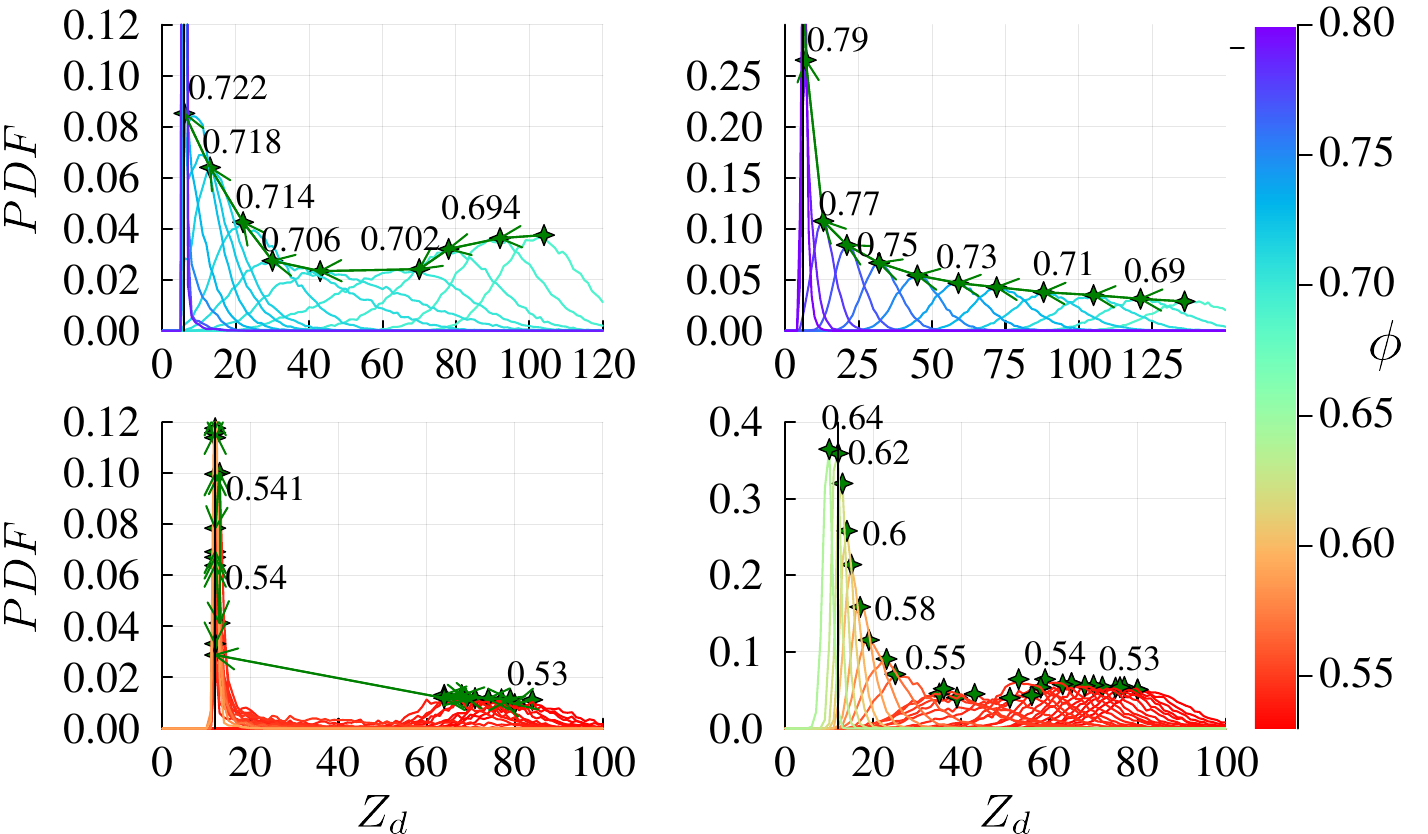}

    \caption{(Color online) PDFs of the number of dynamic neighbors are shown for several packing fractions, with each curve corresponding to a specific value as labeled: (top-left) Monodisperse disks, (top-right) Binary mixtures of hard disks, (bottom-left) Equilibrated hard spheres, (bottom-right) Fast compression of hard spheres. The color of each curve is related to the corresponding packing fraction, as indicated by the color scale on the right side of the figure.  
    }
\label{fig:histt3}
\end{figure}

Figure~\ref{fig:histt3} (top-left) displays the probability density functions (PDFs) $\rho_{Z_d}$ of the number of dynamic neighbors for various packing fractions in a monodisperse hard disk fluid.
In general, increasing $\phi$ leads to an increase in $\max(\rho_{Z_d})$. However, an interesting observation is that for $\phi$ values within the range of $[0.7, 0.72]$, there is a notable drop in $\max(\rho_{Z_d})$. Remarkably, this decrease in $\max(\rho_{Z_d})$ corresponds to the packing fractions associated with the hexatic phase~\cite{engel2013hard, bernard2011two, jaster2004hexatic}.

A similar effect can be observed in Figure \ref{fig:histt3} (bottom-left), where the fluid-crystal transitions of 3D hard spheres are depicted. In this case, a clear jump in the average number of neighbors is evident, transitioning from approximately 60 to 12, which corresponds to the density and order-parameter jumps observed during the phase transition. Conversely, in the case of the 2D binary mixture and the 3D cases with a fast compression rate, crystallization is frustrated and the maximum of the probability distributions increases monotonously with the packing fraction, as shown in the right panels of figure \ref{fig:histt3}.

\begin{figure}
   \centering
    \includegraphics[width=245pt]{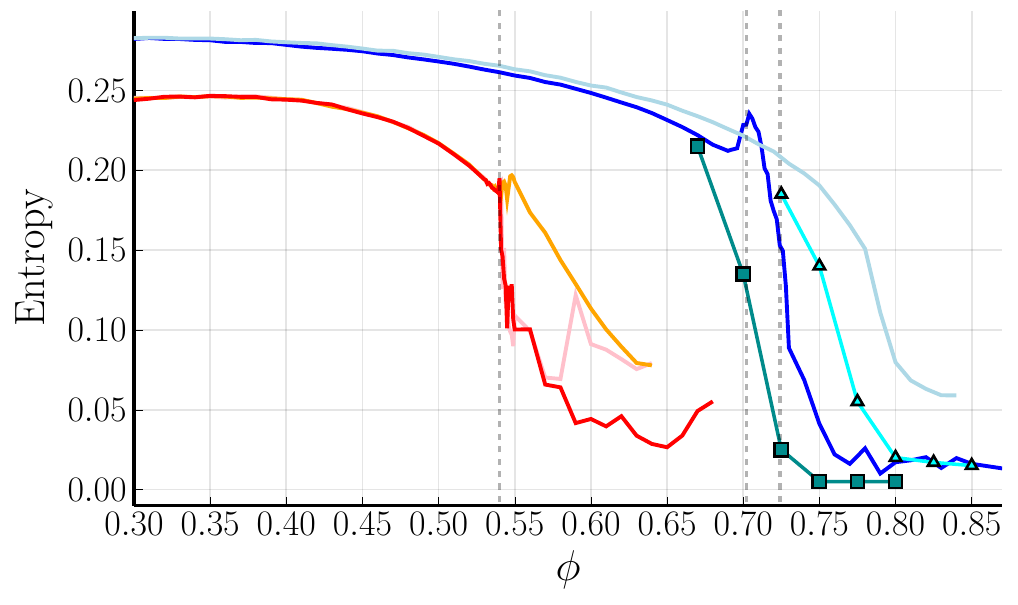}
    \caption{(Color online) Shannon entropy of the $Z_d$ PDFs for the 2D and 3D systems. The blue lines represent the 2D cases, while the red, orange, and pink lines correspond to the 3D cases. Additionally, we have included the data from \cite{donev2007configurational} for the configurational entropy of monodisperse systems (shown in green lines and squares) and the binary mixture of 2D systems (shown in cyan lines and triangles). The vertical dashed lines correspond to specific values of $\phi$, namely $\phi =$ 0.54, 0.702, and 0.724. }
\label{fig:entropy}
\end{figure}

We calculated the Shannon entropy $S = -\sum_{Z_d} p(Z_d) \log(p(Z_d))$ to compare it with the configurational entropy computed elsewhere~\cite{donev2007configurational}, as depicted in Figure \ref{fig:entropy}. The blueish lines represent the 2D systems, while the red line corresponds to a 3D system with a slow compression rate (reaching equilibrium). The light and dark blue curves correspond to the monodisperse and the binary mixture, respectively. The green lines with square symbols and blue lines with triangles correspond to the configurational entropy as reported by Donev et al.~\cite{donev2007configurational}. The orange and pink lines correspond to fast compression rates, with short and large sampling time windows, respectively.  It is worth mentioning that at least another version of Shannon entropy has been compare with the configurational entropy, observing an abrupt drop during the phase transition \cite{walraven2020entropy}. 

The comparison between the configurational entropy (green curve and squares) and the Shannon entropy of the $Z_d$ probability density distributions reveal a consistent match across the entire range of $\phi$, except for the region corresponding to the appearance of the hexatic phase. This correlation suggests that the Shannon entropy of the $Z_d$ distributions is indeed closely related to the configurational entropy.  However, the dark blue curve exhibits a distinct behavior with an entropy peak followed by a sudden decrease, which contrasts with the green curve (note that the disagreement relies only on a single data point). 

The entropy peak observed in the dark blue curve aligns with the reduction in the maximum probability densities observed in Figure \ref{fig:histt3} (top-left) at $\phi = 0.702$. This increase in entropy is consistent with the total system entropy, which includes both configurational and vibrational contributions and is characteristic of an entropy-driven first-order phase transition. Notably, the Shannon entropy peak of $Z_d$ disagrees with the reported datum at $\phi=0.702$ in the literature. The origin of this difference is not clear to us but we believe it may be attributed to the fluid-hexatic coexistence occurring at $\phi=0.702$, where a portion of the system forms a hexatic phase while the remainder remains fluid.

The hexatic fluid has a lower configurational entropy but also a lower specific volume, resulting in the release of free volume during its formation, which increases fluid entropy \cite{frenkel1994simulation}. 
Consequently, the configurational entropy loss due to hexatic formation is partially compensated by the fluid entropy gain. Additionally, the coexistence of the two phases introduces further entropy associated with the different ways of arranging an inhomogeneous system. This additional source of entropy leads to the broadening of the $Z_d$ distribution, explaining the entropic maximum, but we also expect to largely contribute to the configurational entropy, aligning with the expectations for a first-order phase transition. It is worth noting that the entropy peak is expected to sharpen with increasing system size, indicating a discontinuity in the thermodynamic limit. 

In the 3D case, it is important to note the absence of a peak for the red curve. The peak is expected to be present before the sudden entropy drop, but it is too sharp to be captured in our simulations. This difficulty is also present in the 2D case and may explain the low entropy of the data point at $\phi = 0.7$ from the literature. Typically, the peak should appear as a result of the formation of bimodal probability density distributions at the phase transition, which is a hallmark of the coexistence of both phases. However, our 3D simulation systems are not large enough to capture both phases within a single simulation cell, preventing us from observing the bimodal distributions. As a result, the peak is not observed in the entropy curve for the red curve (slow compression case).  

The plots in Figure \ref{fig:entropy} show significant differences in the paths for the fast compression rates with short and large relaxation times within the range of $\phi$ from $0.54$ to $0.59$ (indicated by vertical lines). This behavior signals that a fast compression leads to a metastable state, which evolves over time with glassy dynamics. Additionally, it is important to highlight that the relaxation time increases as the packing fraction rises, as evidenced by the pink curve above $\phi \approx 0.59$. Beyond this critical value of $\phi$, the orange and pink curves almost coincide, further supporting the findings in~\cite{speedy1997pressure, valeriani2012compact}. This observation indicates that, for $\phi > 0.59$, the dynamics of the system approach a steady state, and the relaxation time diverges as $\phi$ increases. This behavior may be associated with the jamming transition that occurs at higher packing fractions.

\section*{Summary}

For systems with hard-core pair potentials, we proposed the definition of the dynamic neighbors of a particle as those that have a collision with the reference one in a given time window. We have shown that the average of the number of dynamic neighbors is connected to the diffusion coefficient and so, it is a dynamic property. In addition, we have employed dynamic neighbors along with the classic definitions of bond-order parameters, which yield similar results to other implementations such as Voronoi tessellations. Finally, we have shown that the Shannon entropy of the dynamic neighbor probability density distributions is closely related to the configurational entropy, at least for the hard-sphere model in 2D and 3D.      

Indeed, there are practical reasons that make the implementation of dynamic neighbors highly useful in various scenarios. Firstly, its direct use for calculating bond-order parameters is straightforward, simpler than using a Voronoi-based implementation, and yields a more pronounced change in the parameter at the fluid-solid transition. This makes it a convenient and efficient tool for studying phase transitions of hard-body systems.

Secondly, dynamic neighbors allow for the direct study of their number average, from which the parameter $\mu_{Z}$ can be obtained. This parameter serves as a measure of the strength of the entropic bond interaction and shows correlations with the behavior of bond-order parameters at fluid-solid transitions. The dynamic nature of $\mu_{Z}$ makes it applicable to systems with intricate solid structures, such as mixtures, polydisperse, and confined systems, as well as for asymmetrical particles. In such cases, finding an appropriate bond-order parameter could be challenging, but $\mu_{Z}$ offers a robust alternative.

Furthermore, for systems of disks and spheres, the Shannon entropy of the dynamic neighbor probability density distributions can be easily computed. It exhibits a remarkable correlation with the configurational entropy, which is a crucial property in understanding glass transitions and other phase transitions. Using the Shannon entropy of dynamic neighbors as a proxy for the configurational entropy makes it a convenient tool for investigating system properties without the computational burden associated with direct configurational entropy calculations.

In summary, the implementation of dynamic neighbors offers practical advantages in studying phase transitions, understanding entropic bond interactions, and correlating with important thermodynamic properties, making it a valuable and versatile approach in various research contexts.

Finally, we would like to emphasize that the concept of dynamic neighbors can be extended to systems with soft potentials by introducing an effective collision distance, at which the interacting particles would have collided as if they were governed by a hard-particle interaction. This extension would allow us to apply the same approach and analyze the dynamics and properties of soft potential systems in a similar manner. By defining an appropriate effective collision distance, we would still capture the essence of the dynamic neighbor concept and use it to study various properties and transitions in soft potential systems.

\section*{Acknowledgments}

We thank David P. Sanders and Michael Schmiedeberg for their valuable discussions. The authors appreciate the computing platform provided by the Laboratorio de Cómputo de Alto Rendimiento, under the coordination of Departamento de Matemáticas of Facultad de Ciencias, UNAM.  

\section*{Disclosure statement}
No potential conflict of interest was reported by the author(s).

\section*{Funding}
We acknowledge the support received through the project PAPIIT IN113923. GO is grateful for financial support through CONAHCyT project A1-S-9197. ASK was supported by Chinese Academy of Sciences President's International Fellowship Initiative for Visiting Scientist grant No 2023VMB0013

\section*{Notes on contributor(s)}

LAIA and ASK designed the concept and numerical experiments, wrote the code for simulations, and wrote the original draft. LAIA performed the simulations. GO and MLH reviewed and edited the original draft. All authors contributed to discussions, the bibliographical review, and data and theoretical analysis. 

\section*{Notes}

The data that support the findings of this study are openly available on a Zenodo repository \cite{data}. 

\vspace{10pt}

\bibliographystyle{apsrev}
\bibliography{main}

\end{document}